\documentclass{emulateapj}

\newcommand{\myemail}{morgan}

\submitted{To appear in {\it The Astronomical Journal}}

\shortauthors{Morgan et al.}

\begin{document}

\title{RETROCAM: A Versatile Optical Imager for Synoptic Studies}

\author{Christopher W. Morgan, Paul L. Byard, D.L. DePoy, Mark Derwent, Christopher S. Kochanek, J.L. Marshall, Thomas P. O'Brien and Richard W. Pogge}
\affil{Department of Astronomy, The Ohio State University, 140 West 18th Avenue, Columbus, OH 43210}

\email{\myemail, byard, depoy, mderwent, ckochanek, marshall, obrien, pogge@astronomy.ohio-state.edu}


\begin{abstract}
We present RETROCAM, an auxiliary CCD camera that can be rapidly inserted into the optical beam     
of the MDM 2.4m telescope. The speed and ease of reconfiguring the telescope to use the imager and a straightforward user interface permit the camera to be used during the course of other observing programs.     
This in turn encourages RETROCAM's use for a variety of monitoring projects.   
\end{abstract}


\section{Introduction}

In an era of 8 meter-class telescopes, the potential for smaller telescopes to make significant scientific contributions can easily be overlooked. In fact, many institutions are ending longstanding relationships with smaller observatories in favor of fewer numbers of nights at larger telescopes. It is clear, however, that dedicated programs on 1--2m class telescopes can still have enormous scientific impact either by conducting large homogeneous surveys such as the Sloan Digital Sky Survey (SDSS) \citep{sloan} or by creative exploration of the time domain. In particular, large scale variability studies have enormous potential: Galactic microlensing studies such as MACHO \citep{macho2000}, EROS \citep{eros} and OGLE \citep{ogle} have placed constraints on the composition and structure of the galaxy.  Dense time-sampling of gamma-ray burst (GRB) afterglows and supernovae has lead to the likely conclusion that the GRB phenomenon is a by-product of supernova events \citep{matheson2003}. The importance of variable star monitoring and its influence on the cosmic distance scale (e.g., Alcock et al. 1995) is well known. Time delay measurements from monitoring multiply imaged quasars constrain  the Hubble constant and the structure of galaxy halos in the important transition region between the baryon dominated inner regions and the dark matter dominated outer regions \citep{kochaneksaasfee2004}.  Additionally, the uncorrelated variability due to microlensing of the images can be used to constrain the fraction of the transition region surface mass density in stars, the mean mass of these stars and the size and structure of the quasar accretion disk \citep{kochanek2004}. 

Rarely do programs of this type require an entire night of telescope time; in fact, many of these variability studies require less than one hour per night of observing time. Unfortunately, traditional scheduling and time-consuming instrument changes make it virtually impossible to pursue such programs at most observatories. Some observatories have overcome this problem with queue or remote observing, but most observatories have multiple primary instruments, some of which are spectrographs with no imaging capability. Hence, consistent synoptic surveys are difficult to accomplish. At some telescopes, the presence of auxiliary ports and bent foci have permitted relatively simple solutions to this problem. The Auxiliary Port Camera (AUX)\footnote{Additional information about AUX can be found at http://www.ing.iac.es/Astronomy/instruments/aux/index.html.} on the 4.2 meter William Herschel Telescope at La Palma is an excellent example of such a system. 
   
RETROCAM, the RETRactable Optical CAmera for Monitoring, is a solution to this problem for the MDM Observatory 2.4m Hiltner telescope. Its primary advantage is that it can be used concurrently with any of the observatory's spectrographs and all but one of its imagers. We initially undertook this project to build a camera for monitoring gravitationally lensed quasar systems, and we will demonstrate RETROCAM's suitability for this task. We will also show that it is capable of much more than this; the primary purpose for this paper is to discuss how RETROCAM's design and capabilities provide an enormous potential for a variety of exciting scientific programs.  

\begin{figure*}[t]
\epsscale{1.0}
\plotone{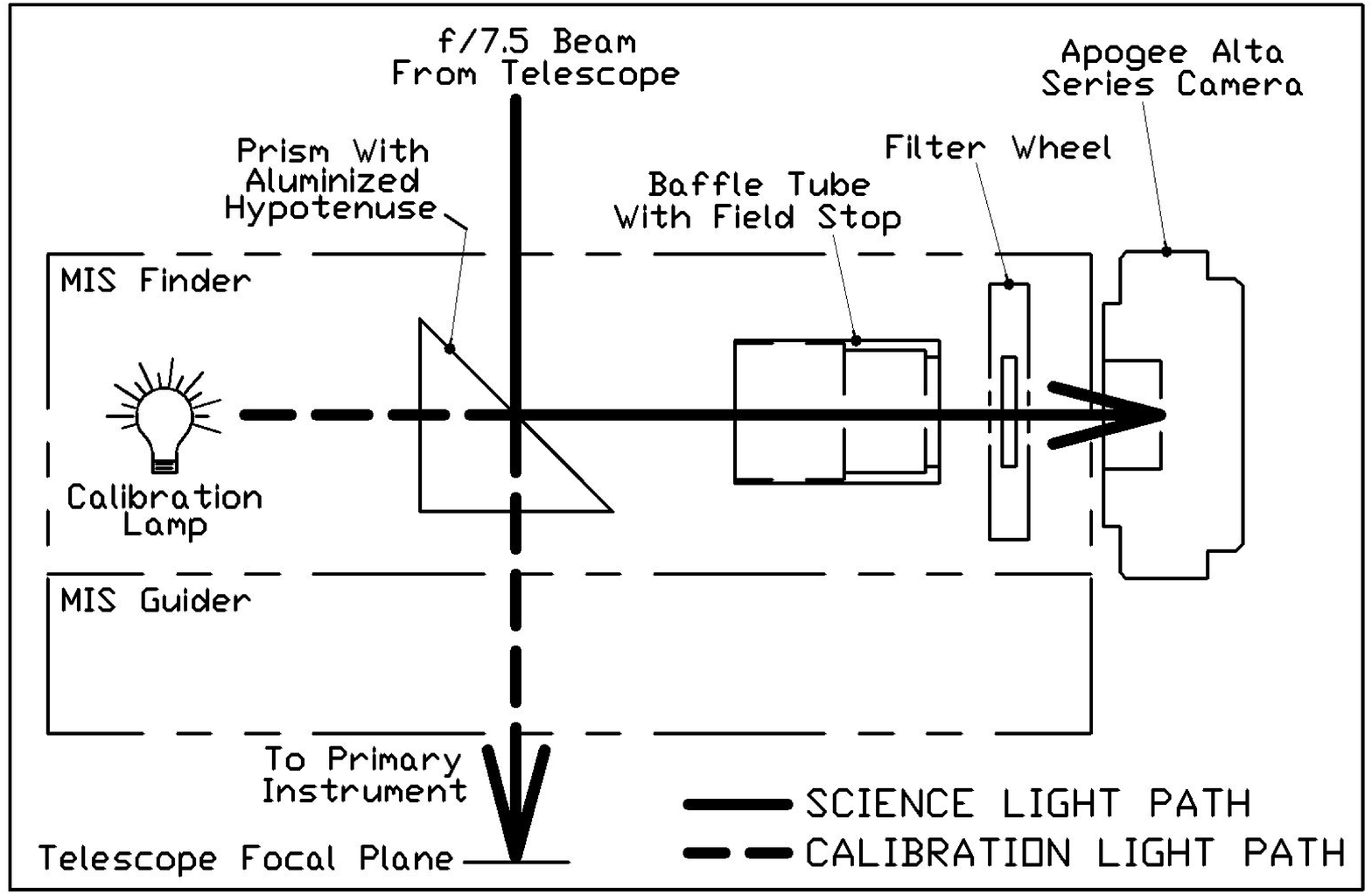}
\caption{A schematic of the RETROCAM and Multiple Instrument System (MIS) optical paths. With the RETROCAM prism inserted, light from the telescope is reflected orthogonally into the camera.  Light from  MIS calibration lamps is directed to the primary instrument below via the prism's internal reflection path. \label{fig1}}
\end{figure*}

\section{RETROCAM Specifications and Configuration}
\subsection{Optomechanical Design}
RETROCAM is installed in the 2.4m Hiltner Telescope's Multiple Instrument System (MIS) at the MDM Observatory on Kitt Peak, Arizona.  The MIS was originally conceived as a multi-function device to provide spectral calibration lamps, an intensified video finder camera, and an x-y stage for the telescope's autoguiding system camera.  Our design preserved the comparison lamp and guider functions, and replaced the antiquated finder camera (a non-functioning intensified vidicon) with a science-quality imager whose quick readout time permits it to also serve as a finder camera, thus maintaining the full original functionality of the MIS and greatly improving its sensitivity.  The layout of the MIS and RETROCAM is shown in Figure 1.  Light from the telescope is sent to RETROCAM by reflection from a reflective optic that travels into the telescope's beam along two rails.  We supress scattered light with a threaded baffle tube and field stop assembly. The light then passes through a filter wheel equipped with four filters  (g,r,i, and z)  from the SDSS set \citep{fukugita96} and one open position.

RETROCAM uses an Apogee Instruments, Inc. Alta E--Series\footnote{http://www.ccd.com/alta.html} CCD camera with the Kodak KAF-1001E-2 CCD array.  The CCD has 24$\micron$ pixels in a 1024x1024 array. When installed on the 2.4m Hiltner Telescope with the \textit{f}/7.5 secondary mirror, the pixel scale is 0.276 arcsec pixel$^{-1}$. The array reads out in $\sim6$ sec with 21 e$^{-}$ pixel$^{-1}$ readout noise at a gain of 7.6 e$^{-}$ DN$^{-1}$. The camera employs a two-stage Peltier Thermoelectric Cooler (TEC) capable of maintaining a differential temperature of $45^{\circ}$C with only ambient air cooling.  The detector is normally maintained at $-20.0\pm1^{\circ}$C and operates with a dark current of 1.5 e$^{-}$ sec$^{-1}$ pixel$^{-1}$ at this temperature. At the sky brightness levels typically seen at MDM in the g,r,i and z bands, noise for all but the shortest exposures is dominated by the contribution from the night sky.  
The camera is mounted on a manually adjustable focusing mechanism that is not meant for frequent adjustment; the focus was adjusted during installation to conjugate the focus of the telescope's most commonly used spectrograph.  The observer accomplishes precise focusing on a nightly basis using the telescope's secondary mirror.

As shown in Figures 1 and 2, the ``Finder'' section of the MIS is located above the guider section.  As a result, when the folding optic is in place it casts a shadow on a significant fraction of the guide field.  In order to maintain the ability to guide during RETROCAM exposures, the folding optic was made as small as possible, preserving $\sim$100 arcmin$^{2}$ of available guide field that is vignetted by $<$20\%. The original MIS design used the back of the finder mirror to reflect light from the calibration lamps into the spectrographs below.  In order to preserve this function while removing the need for a bulky mounting frame, we selected a BK7 prism with an aluminized hypotenuse as our folding optic.  The stiffness of the prism allows it to be mounted using the vertical face only. Light from the telescope reflects directly off of the aluminized surface, while light from calibration lamps is sent downward into the primary instrument via the prism's internal reflection path. A weakness of this design is that BK7 attenuates the calibration lamp light by $\sim$40\% at 4000{\AA} compared to the previous design, so longer comparison lamp exposure times are necessary.  We consider this to be of minimal impact, given the relatively short times typically required for comparison lamp exposures. Spectral quality is otherwise unaffected.

\begin{figure*}[t]
\epsscale{1.0}
\plottwo{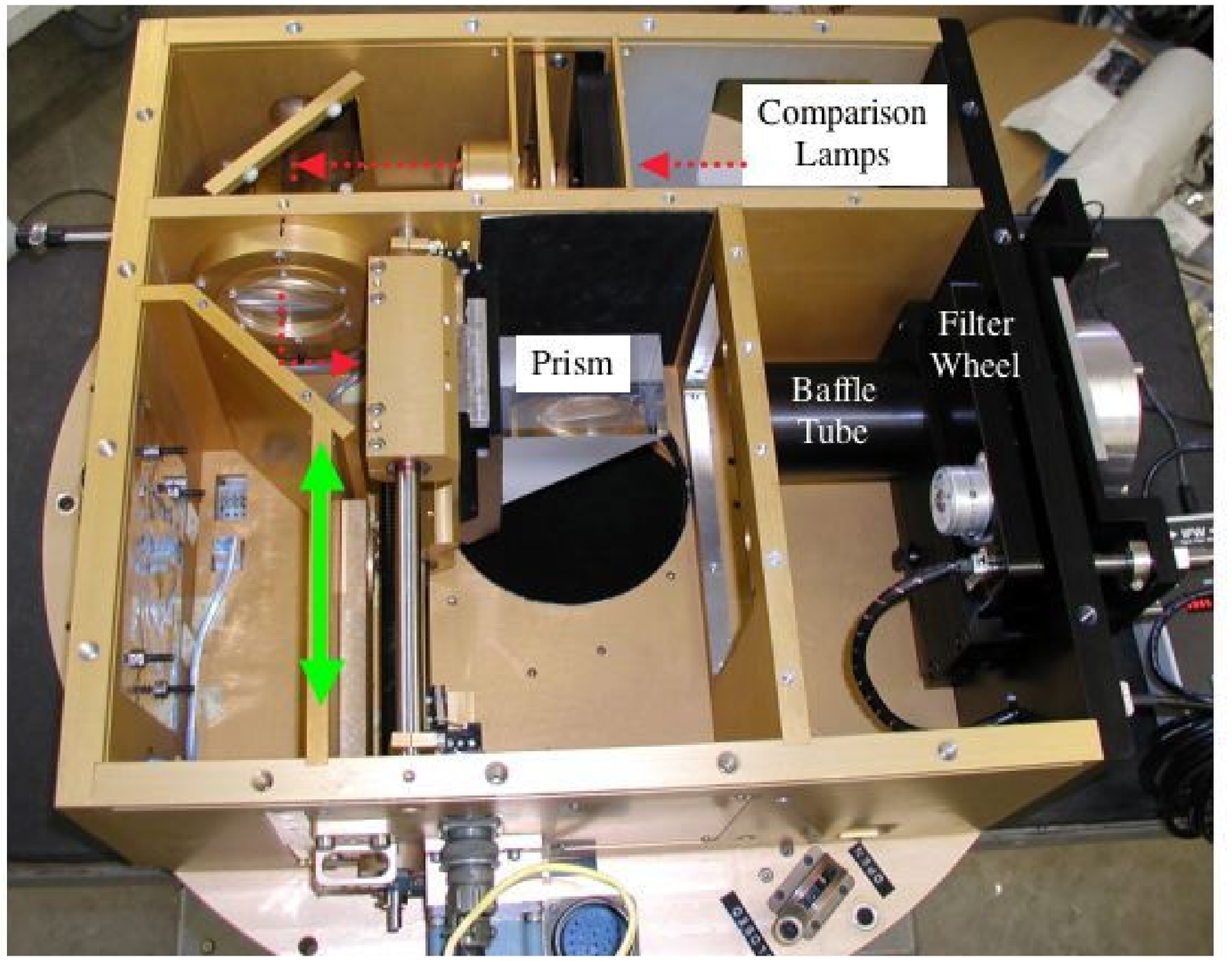}{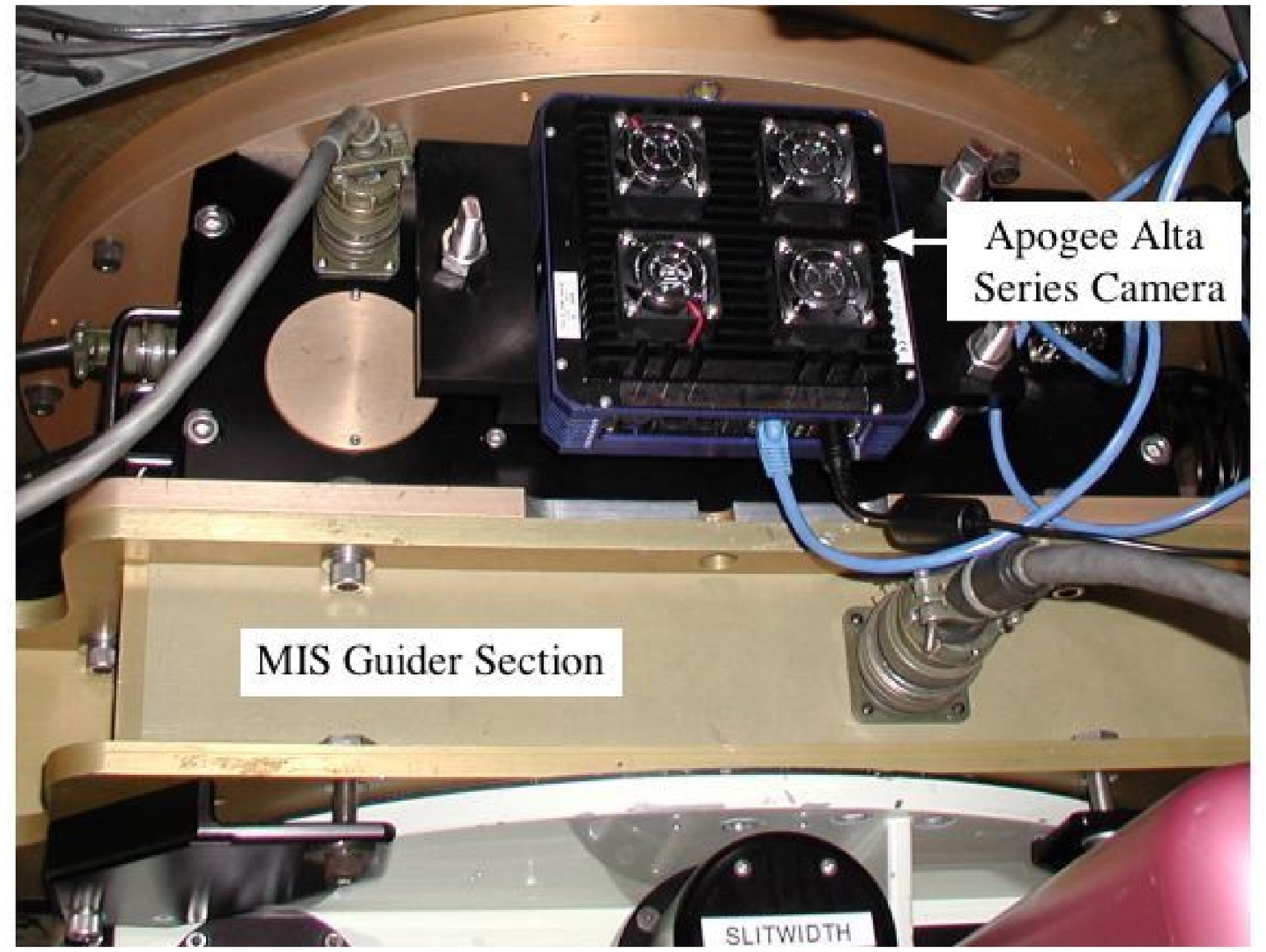}
\caption{Left: Upside-down internal photograph of the MIS finder section during RETROCAM installation. The prism is shown in the fully inserted position. The circular aperture beneath the prism allows light from the telescope to enter the MIS. The prism travels in and out of the optical beam along a set of stainless steel rails.  The baffle tube and filter wheel assembly are fastened to the side wall of the MIS (right side of photo). Compartments in the top of the image house spectral comparison lamps and optics. The dotted red line indicates the comparison lamp light path.  The green arrow indicates prism's direction of motion. Right: Photograph of RETROCAM following installation on the 2.4m Hiltner telescope.  The MIS guider is contained in the gold anodized section of the MIS located beneath the camera.  The white painted structure at the bottom of the photograph is the Boller and Chivens CCD Spectrograph (CCDS).\label{fig2}}
\end{figure*}

RETROCAM's relatively small field of view greatly simplified the overall optical design. No field flattener or off-axis correction lenses were necessary to produce a design Strehl Ratio $>$70\% across the entire field of view.  The current Kodak detector is a square array, and is larger than the design field of view in the East-West direction.  The unvignetted region of the detector is 714x1024 array of $22.5\micron$ pixels, corresponding to a 3.29x4.72 arcmin field of view. The current detector is a proxy for the design model, an Apogee camera employing the E2V CCD55-20 thinned and backside illuminated CCD.  We will install this new detector once it becomes available from the manufacturer. The E2V CCD will provide a 30\% improvement in quantum efficiency, and is properly sized to avoid the vignetting which occurs at the edges of the larger Kodak array. 

\subsection{Data Acquisition and Instrument Control Software}

The Apogee Alta E-Series CCD camera uses an embedded 100-baseT network interface, which is connected directly to the MDM mountain network via a fiber-optic ethernet repeater mounted on the telescope.  The filter wheel we use is an Optec, Inc.\footnote{http://www.optecinc.com/} Intelligent Filter Wheel (IFW).  This filter wheel holds up to five 2-inch diameter filters, and can be operated remotely via an RS-232 serial port using a simple command language provided by Optec.  We access this interface via a Comtrol\footnote{http://www.comtrol.com/} DeviceMaster RTS 4-port network serial port server mounted directly on the MIS box, making the IFW a network-addressable device.  We connect this to the mountain network using the same fiber repeater line that services the Alta CCD camera.

The data acquisition software for the Alta camera runs on a Linux workstation located in the 2.4-m control room.  The user interface is a custom Tcl/Tk application that is based on a slightly modified version of Apogee drivers for Linux, using code provided by Dave Mills\footnote{Version 1.1. See http://sourceforge.net/projects/apogee-driver/ for the most recent version of Mills' driver.}.  Our graphical user interface (GUI) is based on Mills' original gui.tcl script, but substantially modified to incorporate remote control of the Optec IFW filter wheel via the network port server, and to provide other RETROCAM-specific functions (e.g., TEC control).  Data from RETROCAM are written to the hard drive on the Linux workstation in standard FITS format.  The data acquisition system also provides a simple TCP-socket interface for remote command operation from other hosts.  A screenshot of the beta-release version of the RETROCAM data acquisition GUI is shown in Figure 3.  This version of the system is being used during the initial deployment phase, and will be updated based on user feedback from actual use at the telescope.  

The RETROCAM data-acquisition system queries the 2.4m Telescope Control System for the relevant pointing information, and derives its UT time/date information from the data-acquisition computer's clock, which is synchronized with the Kitt Peak mountain network time server. Future versions of the RETROCAM software will be capable of sending commands such as focus offsets to the Telescope Control System.  This remote command capability will permit a high degree of automation which will in turn improve the efficiency of observations with RETROCAM. Inclusions of these functions must await the upgrade of the MDM computer system currently planned for the summer of 2005.

\begin{figure*}[t]
\epsscale{1.0}
\plotone{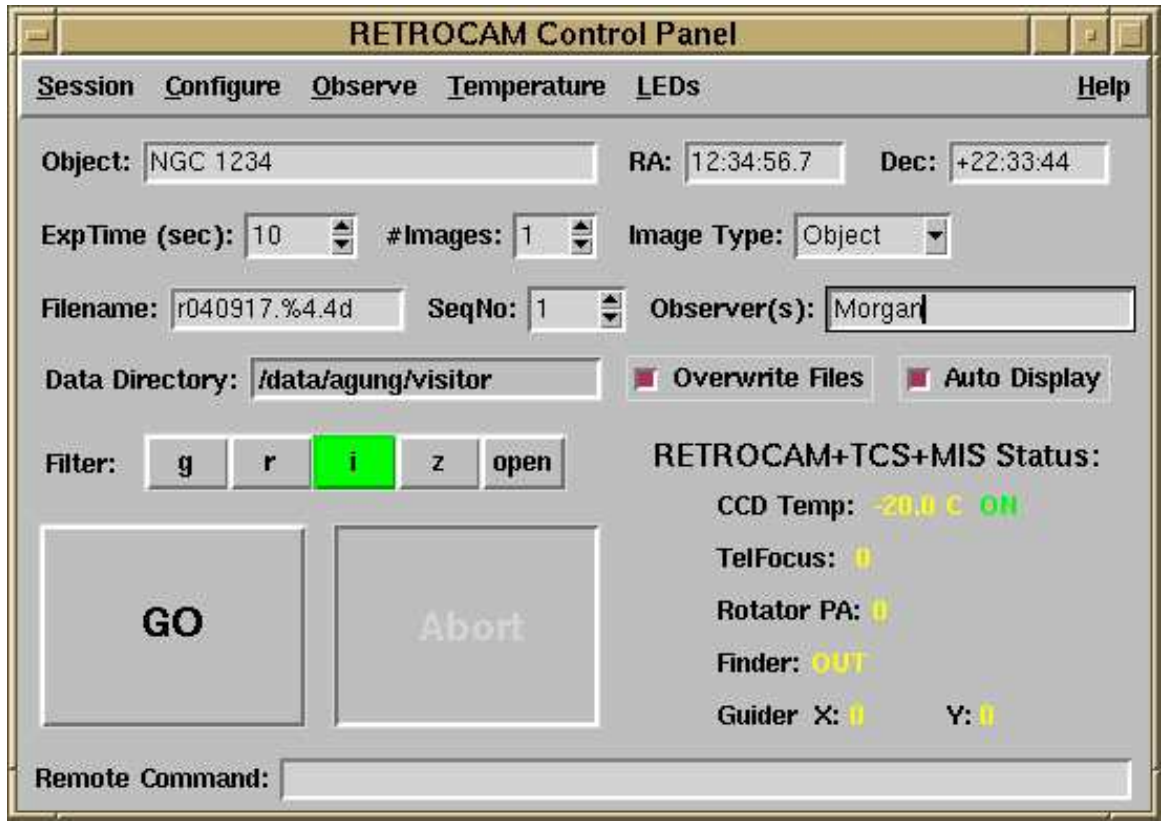}
\caption{The RETROCAM Graphical User Interface (GUI).\label{fig3}}
\end{figure*}

\section{Performance and Usage}
\subsection{Commissioning and Testing}
RETROCAM was installed in August of 2004 and commissioned in September of 2004. We measured a total system (RETROCAM + telescope optics) throughput of 54\% in the r band using the current Kodak CCD. In the r-band, the camera will register $3.11\times10^5$ e$^{-}$ sec$^{-1}$ from an $r=12.0$ magnitude star. Once the front-side illuminated Kodak CCD is replaced by the thinned, back-illuminated E2V CCD, we expect that the efficiency will improve by a factor of $\sim$1.4. 

We evaluated the flatness of the RETROCAM optical system (folding optic + filter + CCD) by measuring the system's response to the flux from a SDSS standard star \citep{smith2002} in a uniform 5x6 grid on the CCD. First, dark current and bias frames were subtracted from each exposure.  Next, we measured the number of detected photons in each image using the Image Reduction and Analysis Facility (IRAF) \textit{qphot} aperture photometry routine. We divided the standard deviation of this sample by the mean and found that $\sigma_{sample}/\mu_{sample} = 0.0057$.  Hence, RETROCAM is flat to within 0.6\% across the entire field of view without application of a flat field correction. Now, the expected value of $\sigma_{sample}/\mu_{sample}$ due to photon flux Poisson noise is $5.7\times10^{-4}$, which is an order of magnitude smaller than what we find. However, it is unlikely that the common procedure of dividing by a normalized flat field (sky or dome flat) will help us approach this theoretical flatness limit. For example, to reach within a factor of three of the theoretical limit, a flat field would need a S/N $\approx600$ and would need to have no systematic (scattered light, etc.) effects greater than $\sim0.1\%$. Obtaining a flat field with such a high S/N and without systematic errors is quite difficult in practice.  As a result, applying a standard flat field correction to RETROCAM images will likely introduce more error than it will remove.

We tested this flat field hypothesis by generating a night sky flat using median combination of a large number of equal length exposures. We normalized this flat and applied it to the SDSS standard star images. We then re-performed the measurement described above and found that $\sigma_{sample}/\mu_{sample}$ increased to 0.011, an increase of nearly a factor of two. This effect is demonstrated graphically in Figure 4. Despite the relatively small number of data points (30), the histogram of the measurements before flattening is obviously narrower than the histogram of the flux measurements following application of the flat field.   We assert that the remarkable flatness of the RETROCAM system is yet another quality which makes it an ideal detector for synoptic studies, since time-consuming flat field corrections will be unnecessary for most projects.

\begin{figure}[t]
\epsscale{1.0}
\plotone{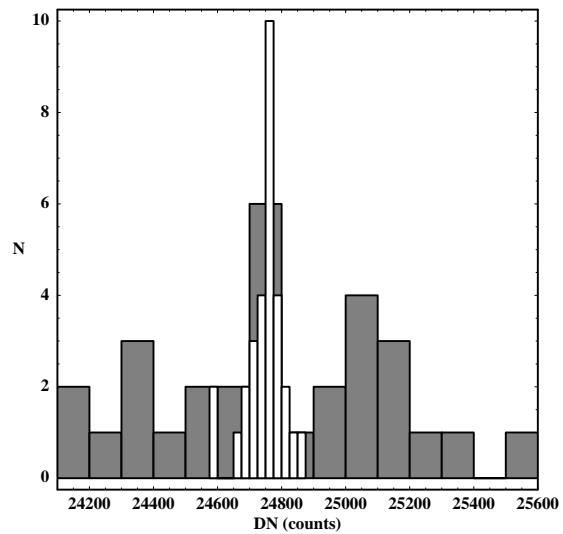}
\caption{A histogram of photometric measurements of the SDSS standard star SA 113 339. Measurements were made at locations in a uniform 5x6 grid on the detector.  White bars: the ensemble of aperture photometric measurements made prior to flat field correction.  Gray bars: The ensemble of aperture photometric measurements following flat field correction. \label{fig4}}
\end{figure}

In the standard RETROCAM observing procedure, a primary observer uses the telescope with the instrument of their choice for the majority of the night.  At a mutually convenient time, the on-site observer takes 15 minutes to reconfigure the telescope and collect data for a dedicated RETROCAM monitoring project, then resumes where they left off with very little time lost to their own program.  None of the settings of the primary instrument have been changed, allowing a rapid return to the primary observing plan.  One complication to this simple plan is that not all primary instruments at the MDM 2.4m are parfocal.
RETROCAM was installed and adjusted to be parfocal with the Boller and Chivens CCD spectrograph\footnote{More information about CCDS can be found at http://www.astronomy.ohio-state.edu/MDM/CCDS/.}.   We have measured the secondary mirror focus offsets between RETROCAM and the other primary instruments at the observatory.  A table of these offsets is provided to the observer to improve the efficiency of the focusing procedure. In an attempt to optimize the efficiency and accuracy of the lensed quasar monitoring project, future improvements to the RETROCAM control software will automate the insertion of the prism, focus offsets, guide star selection, image capture and system restoration. 

\subsection{Sample Results}
An example of RETROCAM's high image quality is shown in Figure 5 - an image of NGC 7714/15 taken in the r band on 2004 August 25.  The image of NGC 7714/15 ($V=13.3$ mag) was taken under 1.0 arcsec seeing conditions. Figure 6 shows an r band image of the gravitationally lensed quasar system Q01420--100 taken 2004 October 2 under 1.1 arcsec seeing conditions. The images are separated by 2.4 arcsec. The brighter image has $R\approx16.6$ mag and the fainter image has $R\approx18.7$ mag. Diffuse light from the lens galaxy can be seen between the two images. 

\begin{figure}[t]
\epsscale{1.0}
\plotone{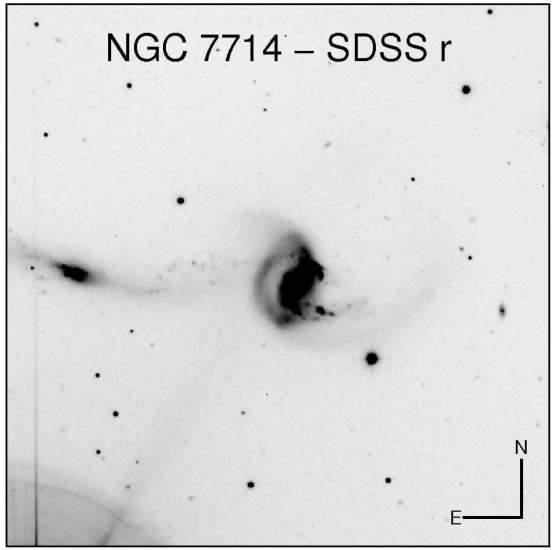}
\caption{An SDSS r band image of NGC 7714 and NGC 7715 taken on 25 August 2004. Orientation axes are 30 arcsec in length. \label{fig5}}
\end{figure}

\begin{figure}[t]
\epsscale{1.0}
\plotone{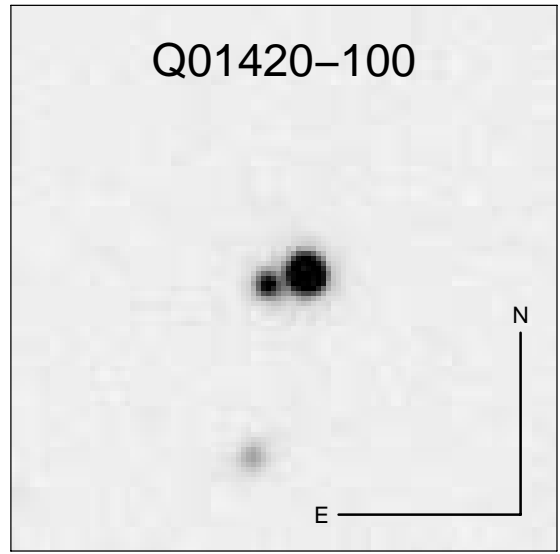}
\caption{An image of the gravitationally lensed quasar system Q01420--100 taken with RETROCAM on 2 October 2004.  The two images are separated by 2.4 arcsec. Orientation axes are 10 arcsec in length. Some diffuse light from the lens galaxy can be seen between the two images. This image was taken under 1.1 arcsec seeing conditions.\label{fig6}}
\end{figure}

\section{Conclusions and Discussion} 
This paper describes the RETROCAM instrument, and demonstrates that it is fully capable of executing the function for which it was designed, but we believe that it will reap returns far beyond lensed quasar monitoring. RETROCAM represents a significant improvement in the capabilities of the MDM observatory in the following ways: (1) It provides the capability to take science quality images and spectra of the same object on the same night. (2) It provides an imager which is nearly always available for targets of opportunity - supernovae, gamma ray bursts, microlensing events, etc.  (3) Its field of view and performance are comparable to most of the direct imagers at the observatory. For many projects, RETROCAM could be used instead of these imagers, reducing the frequency of primary instrument changes and freeing the observatory staff for other activities. RETROCAM will not completely replace these other imagers since it cannot be easily outfitted with observer-supplied filters. 

RETROCAM is currently in use for several variability studies.  C. Morgan and Kochanek are monitoring a sample of gravitationally lensed quasar systems.  A. Gould is using RETROCAM to collect data for the {\AA}ngstrom Project\footnote{More information on the {\AA}ngstrom Project can be found at http://www.astro.livjm.ac.uk/Angstrom/.}, a microlensing study of the M31 bulge. We expect that other members of the MDM consortium will begin other programs soon. RETROCAM represents an important step forward in the design and operational paradigms of the MDM Observatory.  Its design could easily be reproduced at other observatories for relatively low cost.  The availability of commercial ``off the shelf'' high quality CCD cameras with network capability has greatly simplified this task.

\acknowledgments

We appreciate Bob Barr's assistance during RETROCAM installation and testing. We are very grateful to Dave Mills of The Random Factory in Tucson, Arizona who provided us with a pre-release copy of his Linux drivers for Apogee cameras, and for graciously answering our many questions.  Having his code to serve as a base greatly accelerated the development of the RETROCAM data acquisition system, enabling us to get onto the telescope quickly.  We also thank Apogee Instruments, Inc. for providing us with a loan of an Alta E6 CCD camera until our E55 camera is completed. 

Facilities: \facility{MDM(RETROCAM)}.

\end{document}